\documentclass[3p]{elsarticle}
\usepackage[T1]{fontenc}
\usepackage[latin9]{inputenc}
\usepackage{babel}
\usepackage{graphicx}
\usepackage{natbib}
\begin{document}
\begin{frontmatter}
\title{Implications of Delay in Compulsory Mask Wearing - A \textit{What-if} Analysis}

\author[sutd]{Brandon Tay Kaiheng}
\ead{brandontay6@gmail.com}
\author[sutd]{Carvalho Andrea Roby}
\ead{andwea3@gmail.com}
\author[sutd]{Jodi Wu Wenjiang}
\ead{jodiwwj@gmail.com}
\author[smt]{Da Yang Tan\corref{cor1}}
\ead{dayang_tan@sutd.edu.sg}

\cortext[cor1]{Corresponding author}
\address[sutd]{Singapore University of Technology and Design, 8 Somapah Road, Singapore 487372}
\address[smt]{Science, Mathematics and Technology, Singapore University of Technology and Design, 8 Somapah Road, Singapore 487372}

\begin{abstract}
We investigate the impact of the delay in compulsory
mask wearing on the spread of COVID-19 in the community, set
in the Singapore context. By using modified
SEIR-based compartmental models, we focus on macroscopic population-level analysis of the relationships
between the delay in compulsory
mask wearing and the maximum infection,
through a series of scenario-based analysis. Our analysis suggests that
collective masking can meaningfully reduce the transmission of COVID-19
in the community, but only if implemented within a critical time window
of approximately before 80 - 100 days delay after the first infection
is detected, coupled with strict enforcement to ensure compliance
throughout the duration. We also identify what is called a \textit{point
of no return}, a delay threshold of about 100 days
that results in masking enforcement having little significant impact
on the Maximum Infected Values.
\end{abstract}

\end{frontmatter}
\section{Introduction}

In the light of the COVID-19 pandemic, the Singapore government declared
with effect as of 14 April 2020, an enforcement of compulsory mask
wearing in public spaces, 89 days after the first case of COVID-19
was detected in Singapore. 

The usage of masks is known to effectively decrease the infection
rate. It has shown success in limiting community spread of SARS 2003 \cite{lau2004sars} , and more recently, in Taiwan\textquoteright s
management of COVID-19 \cite{wang2020response}. Recent hypothetical
studies on masking by the states of New York and Washington suggests
a potential prevention of up to 45\% of their projected death rates \cite{eikenberry2020mask}. In this study of potential face-mask usage
for the general public, the authors investigated how public masking
can control the infection, in the context of the USA. However, not
much was mentioned about how the delayed enforcement of such a policy
would affect the infection numbers. In another study on public masking \cite{howard2020face}, the authors studied how factors such as the
filtering capability of different mask materials, as well as sociological
behaviour patterns on masking, would affect the efficacy of this policy.
However, in-depth studies on the impact of delayed mask enforcement
on the spread of COVID-19 in the community is limited.

This paper thus seeks to investigate the relationship between delay in compulsory mask wearing and the Maximum Infected Values, through a series of scenario-based what-if analysis. We would be considering 3 scenarios using 2 compartmental models - the SEIR and the SEIRS model. For the SEIR model, we consider a complete compliance, and gradual noncompliance over time. Using an SEIRS model, we consider a third scenario taking into account time-limited immunity. Results from simulating these 3 scenarios would hopefully shed light on the effects of delaying the usage of masks and how the containment of epidemic could have been more effective in Singapore. 

To model the pandemic from a macroscopic view point, we use modified
versions of the model, which consists of a number of compartments
described by a system of differential equations in Sections \ref{sec:SEIR} and \ref{sec:SEIRS}. To solve these equations, we implemented the model in Python using
the Odeint package. We then modify the Delay of Mask Enforcement
by plotting several data sets and recording the resulting Maximum
Infection Value predicted by our model. This was performed over 3 different
scenarios:

\begin{itemize}

\item Scenario 1: The first scenario considers the most basic case of complete compliance of the masking enforcement, from the day it was enforced throughout its duration. It also does not account for time-limited immunity \cite{kosinski2020influence}, where infected individuals become susceptible again after a period of time. 

\item Scenario 2: The second scenario closely resembles the first, but with the addition of gradual non-compliance of the masking enforcement. The root of noncompliance stems from either lack of medical knowledge, wishful thinking that the pandemic will magically disappear, selfish behaviour of individuals, pandemic fatigue etc. Here, we assume that onset of the noncompliance is triggered by an event, for instance, changes in government policies. As an illustration, the Singapore government announced the Phase 2 of its gradual reopening plan on 18 June 2020 (154 days after the first case in Singapore) where members of the public were allowed to visit shopping malls and dine-in at food establishments, it was observed that a minority of individuals did not comply with the masking regulations. 

\item Scenario 3: The third scenario is almost identical to the first, but now accounts
for time-limited immunity. After 90 days, it has been shown that
a recovered individual may become susceptible to the virus again \cite{ibarrondo2020rapid}. This scenario serves as a comparison
to investigate if the immunity factor would worsen the effects of
the delay of mask enforcement on the maximum infected values. 

\end{itemize}

\section{SEIR Model} \label{sec:SEIR}

The SEIR model consists of several compartments, namely Susceptible,
Exposed, Infected and Removed. We expand the model by further
differentiating the Removed compartment into 2 new Recovered and Death
compartments. The flow between the 5 components are described by the following 5
differential equations: 

\begin{equation} \label{eq:SEIR-S}
\frac{dS(t)}{dt}=-\beta I(t)\frac{S(t)}{N}
\end{equation}

\begin{equation}
\frac{dE(t)}{dt}=\beta I(t)\frac{S(t)}{N}-\delta E(t)
\end{equation}

\begin{equation}
\frac{dI(t)}{dt}=\delta E(t)-(1-\alpha)\gamma I(t)-\alpha\rho I(t)
\end{equation}

\begin{equation} \label{eq:SEIR-R}
\frac{dR(t)}{dt}=(1-\alpha)\gamma I(t)
\end{equation}

\begin{equation}
\frac{dD(t)}{dt}=\alpha\rho I(t)
\end{equation} where $N$ is the total population, $S(t), E(t), I(t), R(t)$ and $D(t)$,
are the number of people susceptible, exposed, infected, recovered
and dead on day $t$. $\beta$ is the expected number of people an infected person infects per day, $\gamma$ is the proportion of recovery per day, $\delta$ is the incubation period, $\alpha$ is the fatality rate due to the infection and $\rho$ is the inverse of the average number of days for an infected person to die if he does not recover.

\section{SEIRS Model} \label{sec:SEIRS}

The SEIRS model is similar to the SEIR model such that it has five components as well - Susceptible, Exposed, Infected, Recovered and Death. However, the SEIRS model takes time-limited immunity into account, whereby recovered individuals are prone to becoming susceptible to the disease again after a period of time. 

In the SEIRS model, Eq. \ref{eq:SEIR-S} and \ref{eq:SEIR-R} are then modified to

\begin{equation}
	\frac{dS(t)}{dt}=-\beta I(t)\frac{S(t)}{N}+\epsilon R(t)
\end{equation}

\begin{equation}
	\frac{dR(t)}{dt}=(1-\alpha)\gamma I(t)-\epsilon R(t)
\end{equation} where $\epsilon$ is the rate at which a recovered person becomes susceptible again.




\section{Time-Based Model for Compulsory Mask Wearing}
\begin{figure}
	\centering
	\includegraphics[scale=0.45]{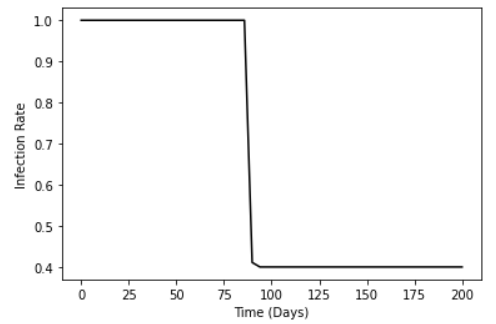}
	\caption{Logistic function to model the infection rate due to compulsory mask wearing $m(t)$ for Scenarios 1 and 3.}
	\label{fig:mtbasic}
\end{figure}

Masking is found to decrease the infection rate $I(t)$ by reducing the transmission of respiratory droplets between individuals, which in turn reduces the number of individuals an infected person can infect. Kai et al. \cite{kai2020universal} showed that the infection rate can be reduced by 60\% when universal masking is enforced. We thus model our infection rate as time-dependent $\beta \rightarrow \beta(t)$ and a function of $m(t)$:

\begin{equation}
	\beta(t)=\kappa m(t)
\end{equation} where $\kappa$ is an arbitrary constant and 

\begin{equation} \label{eq:maskbasic}
	m(t)= \frac{\beta_{s}-\beta_{c}}{1+e^{-k(-t+t_0)}}+\beta_{c}
\end{equation} where a modified logistic function is used to model this transition by setting the infection before (i.e., at the \textit{start} of the outbreak) and after the full \textit{compliance} masking enforcement to be $\beta_{s}=1$ and $\beta_{c}=0.4$ respectively. $t_0$ is the number of days after the first case where masking wearing is enforced. Figure \ref{fig:mtbasic} shows an example of such logistic function, where for the case of Singapore, we set $t_0=89$, since the policy of compulsory mask wearing was implemented 89 days after the first case was uncovered. This model is used in both Scenarios 1 and 3.

To model the gradual noncompliance of universal masking, we make a modification to Eq. \ref{eq:maskbasic}.

\begin{equation}
	m(t)= \frac{\beta_{s}-\beta_{c}}{1+e^{-k_1(-t+t_0)}}+\frac{\beta_{c}-\beta_{nc}}{1+e^{-k_2(-t+t_1)}}+\beta_{nc}
\end{equation}

\begin{figure}
	\centering
	\includegraphics[scale=0.45]{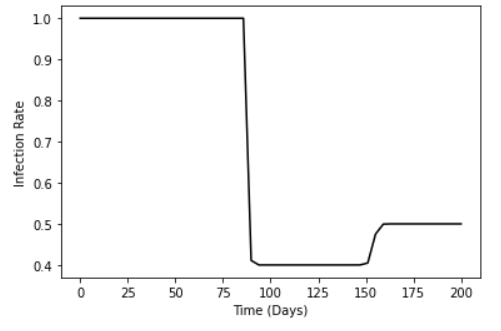}
	\caption{Logistic function to model the infection rate due to compulsory mask wearing $m(t)$ for Scenario 2, where one takes into account the noncompliance.}
	\label{fig:mtwithnc}
\end{figure}

As individuals began to engage in \textit{noncompliance} out of complacency and pandemic fatigue, the infection rate, $\beta_{nc}$, would thus increase slightly. Furthermore, compared to quenching of the infection rate due to enforcement of the compulsory mask wearing, the noncompliance will be gradual. This results in a gentler gradient as the infection rate transits from $\beta_{c} \rightarrow \beta_{nc}$. The gradients are tuned by the arbitrary constants $k_1$ and $k_2$, where $k_1 > k_2$. We further assume that the population are able to maintain compliance for a period of time before onset of noncompliance at $t_1$, which may be triggered by an event, for example a change in the government's policy. Figure \ref{fig:mtwithnc} illustrates one such example, where we have $\beta_{nc}=0.5$ and $t_1=154$, taken in context to Singapore's shift from a full lockdown to gradual resumption of everyday activities 154 days after the first case. The infection rate $\beta_{nc}$ is expected to be lower than $\beta_{s}$ as even with the complacency and fatigue, as there is now a greater situational awareness of the severity and the population in general will take a more cautious outlook compared to pre-pandemic days.

\section{Epidemiological Parameters}

To conduct our analysis in the Singapore context, we estimate the epidemiological parameters described in the earlier sections for the Singapore's context:

\begin{itemize}
	\item $\gamma=1/11$: According to the position statement released by the National Centre of Disease Control Singapore \cite{min2020position}, a person remains infectious for up to 11 days after first contracting COVID-19.
	\item $\delta = 1/5$: The SARS-CoV2 virus has an incubation period of about 5 days \cite{lauer2020incubation}.
	\item $\alpha = 0.000064$: Fatality rate is defined as the percentage of deaths among all previously infected individuals. At the time this work was conducted, the number of deaths in Singapore was 26, and the total number of Recovered and Dead compartments was 40,625.
	\item $\rho = 1/9$: As this number varies greatly across different demographics and is highly unpredictable, we are unable to obtain a proper average. Moreover, owing to the low numbers of COVID-19 deaths in Singapore, it would be inaccurate to calculate an average using this small sample size. Thus, for analysis purposes, we set it at 9 days \cite{froese2020}.
	
\end{itemize} 

\section{Results and Discussion}
\subsection{Dynamics of Delay of Compulsory Mask Wearing}
\begin{figure*}
	\centering
	\includegraphics[scale=0.48]{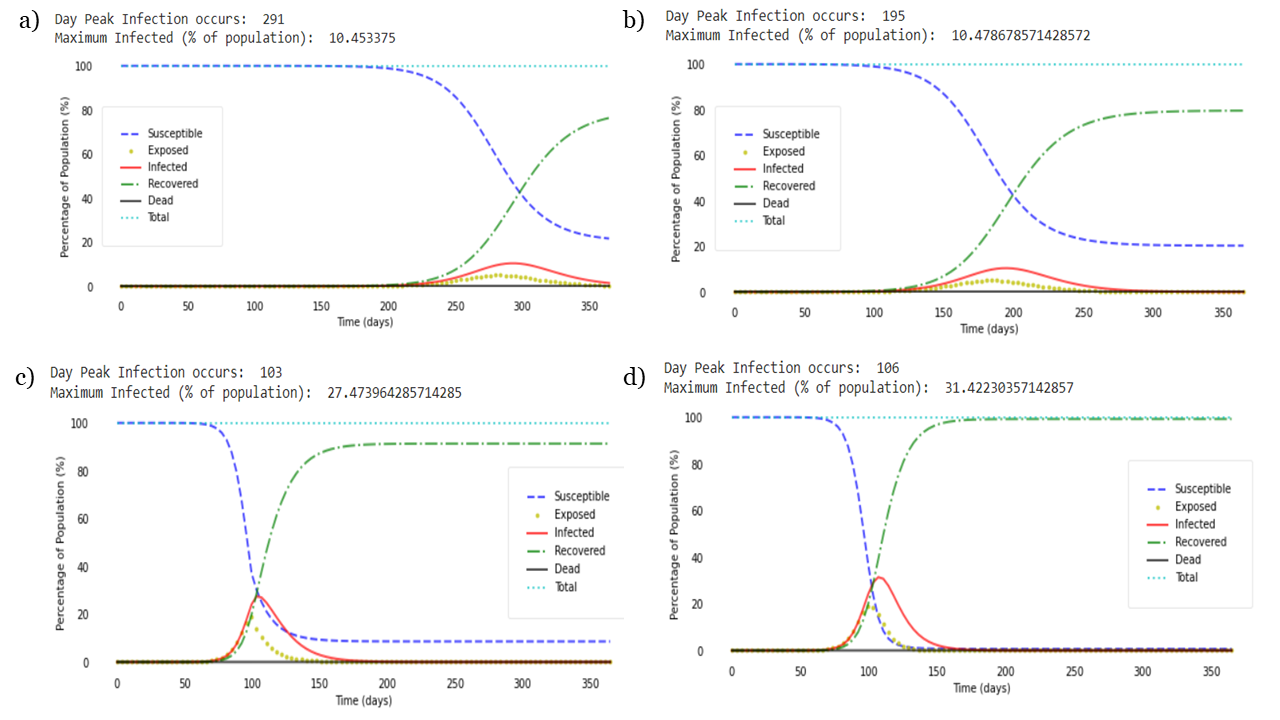}
	\caption{SEIR Plots for delays of (a) 0 days, (b) 50 days, (c) 100 days. (d) corresponds to the control case where mask wearing is not enforced}
	\label{fig:maskingwearSEIR}
\end{figure*}

Upon applying the SEIR models to the 3 scenarios described in the above sections, we are able to obtain Maximum Infected Values by simulating different values of delay in mask enforcement. Considering the most basic case of Scenario 1; where there is complete compliance throughout the duration of mask enforcement, and the absence of time-limited immunity, Figure \ref{fig:maskingwearSEIR} shows the cases where mask wearing were to be enforced (a) on the day the first case of COVID-19 was detected in Singapore; (b) after a 50 days delay; (c) after a 100 days delay; and (d) not enforced at all. 

From Fig. \ref{fig:maskingwearSEIR}, it can be deduced that earlier enforcement of mask wearing both reduces the Maximum Infected Value as well as increases the number of days taken to reach the Maximum Infected Value. This is reflected by right-ward shift of the maxima of the Infected compartment, as well as a lower maxima value of the infected curve. The infected number increases until it reaches a global maxima, before decreasing as predicted by compartmental models of such form. This global maxima is termed as the Maximum Infected Value.  For the case of (a) 0 days of delay in mask enforcement, the maximum infected value is approximately 10.453\%, and the peak occurs at day 291 after the first case was detected. For the case of a (b) 50 days delay, the maximum infected value is approximately 10.479\%, and the peak occurs on day 195. For the case of a (c) 100 days delay, the maximum infected value is much higher, at approximately 27.474\%, and the peak occurs much earlier, on day 103. This is very close to the case where (d) no public masking is enforced, where the maximum infected value is approximately 31.422\% and the peak occurs on day 106. From a policy point of view, this suggests that early universal masking would indeed be an effective control, not only to reduce the infection, but more critically to flatten the curve so as to not overwhelm the medical resources.

\subsection{Effects of Delay of Mask Enforcement on Maximum Infected Value}
\begin{figure}
 	\centering
 	\includegraphics{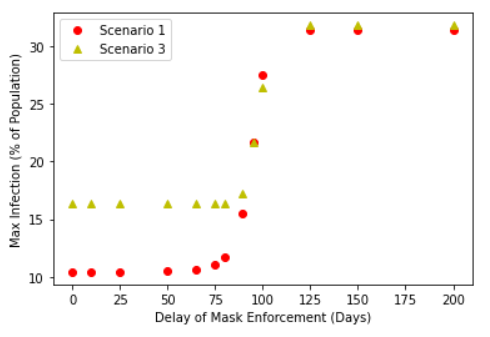}
 	\caption{Maximum infection against delay in mask enforcement for Scenario 2: SEIR with full compliance and Scenario 3: SEIRS with time-limited immunity.}
 	\label{fig:maskvsdelay}
 \end{figure}

 We further investigate this relationship by explicitly considering how the delay in mask enforcement will impact the corresponding Maximum Infected Values. Figure \ref{fig:maskvsdelay} shows the results of different enforcement delay values and their corresponding maximum infected values. Here, we considered 3 scenarios described in the introduction. 
 
Fig. \ref{fig:maskvsdelay} suggests a transition from low maximum infection of about 15\% - 16\% to a high maximum infection (which we dub as the \textit{point of no return}) of about 31\%, with the transition occurring at between a delay of 80 to 100 days. Interestingly, this suggests that once we cross this point of no return of about 100 days, enforcing compulsory mask wearing will have little effect on the Maximum Infected Value. This is because after 100 days, most of the Exposed compartment has already been been infected, therefore this will no longer contribute to any further increase to the Infection compartment. Relating this to the Singapore's context, the compulsory mask wearing was introduced 89 days after the first case, hence this suggests that the introduction of compulsory mask wearing was timely to prevent such point of no return. 

Crucially, one should note that such point of no return appears to be independent of the choice of our scenarios and in all three cases, the transition takes place at about the same 80 to 100 days of delay. The peak infection beyond 100 days of delay also yield similar results in all three cases. One would have naively anticipate that the noncompliance in Scenario 2 resulting in a larger infection rate and the backflow of Recovered to Susceptible compartment after 90 days of time-limited immunity would have contributed positively to the Maximum Infected Values. While this is indeed the case for early intervention, beyond the transition point, the Infected compartment is already on track in reaching its maximum (see Fig. \ref{fig:maskingwearSEIR}(d)), any intervention will no longer be able to significantly change its trajectory, nor effectively reduce the maximum. This suggests the importance of early intervention by policy makers and governments, where this potential window to take action is about 3 months based on our analysis.
 \begin{figure}
	\centering
	\includegraphics{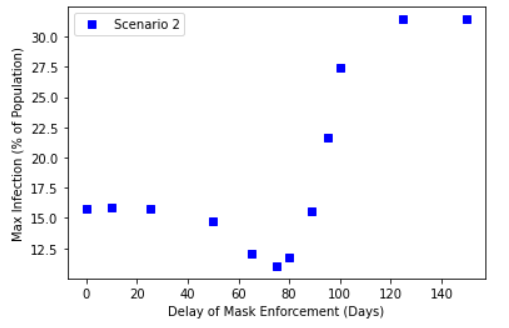}
	\caption{Maximum infection against delay in mask enforcement for Scenario 2: SEIR with gradual noncompliance.}
	\label{fig:maskvsdelaycase2}
\end{figure}
We further note that in Scenario 2 (see Fig. \ref{fig:maskvsdelaycase2}), the results suggest that earlier enforcement of mask wearing leads to higher than expected Maximum Infected Values (compared to Scenario 1). This is apparent for delays in mask enforcement under approximately 50 days. It is likely that for such cases of early enforcement, the Susceptible population remains very high throughout. Consequently, when the agents begin to flout the rules after $t_1=154$ days, the combination of both larger Susceptible compartment pool and higher $\beta_{nc}$ results in a greater amount of infection. In other words, the susceptible population must be sufficiently reduced before the implementation of the compulsory mask wearing for it to be effective in reducing the Maximum Infected Value. In practice, though not considered in this work, the susceptible population may be reduced or removed through other means, e.g. social distancing or lockdown.



\subsection{Limitations of the Results}

We caution against \textit{exact} quantitative predictions of the development of pandemic that are dependent on a multitude of factors other than universal mask wearing that we have considered here. What we have considered here is the \textit{what-if} analysis of the effects of implementation of compulsory mask wearing on the dynamics of the pandemic. While the parameters chosen were based on estimates in Singapore's context, one should note that at the time of this writing, these estimates may change as we continue to deepen our understanding of the virus spread. 

Instead, this work should be understood \textit{qualitatively} as a basis for the discussion of the general trends. We further note that, at the time of writing, the actual real-world figures for Maximum Infected Values fall below the ones discussed in this paper. This is due to the fact that we have only examined the policy of compulsory mask wearing in isolation, with minimum consideration to other factors. We expect that any countries or territories that are actively fighting the virus spread to implement an array of varying measures, all of which would collectively reduce the overall spread of the virus. 

\section{Concluding Remarks}
At present, compulsory mask wearing has been widely accepted as a means of controlling COVID-19 infection by reducing the infection rate through aerosol means. With this study, we hope to shed some light on whether the delay in enforcement of compulsory mask wearing will have detrimental effects on infection control. Based on our results, it appears that a delay of 100 days and above would result in a \textit{point of no return}, where enforcement of public masking would result in little effect on controlling the Maximum Infected Value. We considered 3 scenarios over 2 varied mathematical models. This result is consistent regardless of the level of compliance of the population, or the presence of time-limited immunity. Yet, if implemented early, our model shows that the Maximum Infected Values can be kept relatively low and under control, even after accounting for time-limited immunity and some level of noncompliance. 

\section{Acknowledgements}
The authors would like to acknowledge the support from the Singapore University of Technology and Design's Undergraduate Research Opportunities Programme (UROP).

\bibliographystyle{elsarticle-num}
\bibliography{maskwearingbib.bib}
\end{document}